\documentclass[aps, prl,twocolumn,superscriptaddress,notitlepage,nofootinbib, 10pt]{revtex4-2}

\usepackage[utf8]{inputenc}
\usepackage[T1]{fontenc}
\usepackage{graphicx}
\usepackage{orcidlink}
\usepackage{amsmath,amsfonts,amsthm,amssymb} 
\usepackage{dsfont}
\usepackage{mathtools}
\usepackage{enumitem}
\usepackage{times}
\usepackage{bm}
\usepackage[dvipsnames]{xcolor}
\usepackage[most]{tcolorbox}
\usepackage{hyperref}
\hypersetup{
    colorlinks=true,
    linkcolor=Green, 
    citecolor=Green,
    filecolor=Green,      
    urlcolor=Green,
    pdftitle={Structured Non-Locality and Emergent Locality in Cavity-QED Many-Body Dynamics},
    pdfpagemode=FullScreen,
    }

\renewcommand{\i}{\mathrm{i}}
\newcommand{\e}{\mathrm{e}} 
\newcommand{\bra}[1]{\langle #1 |}
\newcommand{\ket}[1]{| #1 \rangle}
\newcommand{\inprod}[2]{\langle #1 | #2 \rangle} 
\newcommand{\outprod}[2]{\ket{#1} \hspace{-1pt} \bra{#2}} 

\begin{document}

\title{Structured Non-Locality and Emergent Locality in Cavity-QED Many-Body Dynamics}

\author{Mark A. Oehlgrien \orcidlink{0009-0002-9672-4435}}
\affiliation{ICFO-Institut de Ciencies Fotoniques, The Barcelona Institute of Science and Technology, 08860 Castelldefels (Barcelona), Spain.}

\author{Błażej Jaworowski \orcidlink{0000-0002-0907-7405}}
\affiliation{ICFO-Institut de Ciencies Fotoniques, The Barcelona Institute of Science and Technology, 08860 Castelldefels (Barcelona), Spain.}

\author{Darrick E. Chang \orcidlink{0000-0002-5426-7339}}
\affiliation{ICFO-Institut de Ciencies Fotoniques, The Barcelona Institute of Science and Technology, 08860 Castelldefels (Barcelona), Spain.}
\affiliation{ICREA-Instituci\'o Catalana de Recerca i Estudis Avan\c{c}ats, 08015 Barcelona, Spain.}

\author{Charlie-Ray Mann \orcidlink{0000-0003-1320-2624}}
\affiliation{ICFO-Institut de Ciencies Fotoniques, The Barcelona Institute of Science and Technology, 08860 Castelldefels (Barcelona), Spain.}

\begin{abstract} 
Cavity quantum electrodynamics (QED) modifies many-body systems by combining cavity-mediated collective interactions with microscopic short-range interactions. 
The resulting dynamics lies between the local and fully collective limits, such that neither locality nor collectivity alone provides a complete organizing principle. 
In the clean conceptual limit of dominant collective coupling, we show that the effective dynamics within each energetically isolated subspace is generically controlled by whether that subspace admits a local product-state basis, and identify exceptions imposed by angular-momentum selection rules.
Product-state subspaces generically retain the spatial structure of the microscopic interaction. 
Entangled subspaces instead generically dress local processes with global operators, generating non-local but highly structured dynamics. 
We illustrate this by deriving the corresponding effective Hamiltonians in two representative cavity-QED spin models.
The cavity-isolated subspace thus becomes a resource for generating competing short-range interactions or globally conditioned local processes, opening a class of many-body dynamics without local or fully collective counterparts.
\end{abstract}

\maketitle


\textbf{\textit{Introduction.---}}
Cavity quantum electrodynamics (QED) can reshape and induce phenomena in many-body systems by combining microscopic short-range interactions with cavity-mediated long-range interactions \cite{Mivehvar_AdvPhys_2021_cavity, Schlawin_ApplPhysRev_2022_cavity, Lu_AdvOptPhotonics_2025_cavity}.
This opportunity also creates a conceptual challenge: neither locality nor collectivity fully organizes the resulting dynamics. In purely short-range systems, locality bounds the speed at which information can propagate \cite{Lieb_Robinson_CommMatPhys_1972_the, Hastings_Koma_CommMatPhys_2006_spectral, Nachtergaele_Sims_CommMatPhys_2006_lieb, Chen_RepProgPhys_2023_speed}, allowing general statements about the decay of correlations \cite{Hastings_Koma_CommMatPhys_2006_spectral, Nachtergaele_Sims_CommMatPhys_2006_lieb} and the classification of gapped phases of matter \cite{Hastings_Wen_PRB_2005_quasiadiabatic}. 
In the fully collective limit relevant to many atomic platforms, permutation symmetry removes spatial structure and confines the dynamics to a small fraction of the Hilbert space, often reducing the many-body problem to a few collective degrees of freedom \cite{Dicke_PhysRev_1954_coherence, Defenu_RevModPhys_2023_long}. 
The study of cavity-coupled quantum matter requires new general organizing principles, which help to predict what types of phases and dynamical regimes emerge and which insights are universal rather than model specific.

We seek such organizing principles in the clean conceptual limit realizable in atomic cavity platforms, where cavity-mediated interactions dominate over microscopic short-range terms. 
The dominant collective interaction separates the many-body spectrum into subspaces labeled by global spin quantum numbers. 
Perturbative short-range interactions must then be projected into each subspace.
Despite the global origin of the constraint, the effective Hamiltonian within a subspace can remain local \cite{Santos_PRL_2016_cooperative, Celardo_PRB_2016_schielding, Mann_2025_squeezing} and even support rich many-body phases. 
For example, projecting Ising interactions into the global spin singlet subspace can transform an otherwise classical antiferromagnet into an exotic quantum spin liquid \cite{Mann_2025_squeezing}. 
However, these previous works leave open which structural property determines whether projection preserves locality and what forms of non-locality emerge when it does not.

Here, we show that the effective dynamics within a subspace is generically organized by whether the subspace admits a local product-state basis, with angular-momentum selection rules producing controlled exceptions.
We call a subspace a \textit{product-state subspace} when it admits a complete basis of states that factorize over sites.
In such a subspace, projection generically preserves the spatial structure of the microscopic short-range interaction, while the operator content can change.
By contrast, we call a subspace an \textit{entangled subspace} when it admits no such basis.
Projection into such subspaces generically dresses local processes with global operators, producing non-local but highly structured effective interactions.
We establish this organizing principle in two representative cavity-QED spin Hamiltonians that realize product-state and entangled subspaces, respectively.
We also identify exceptions that arise from angular-momentum selection rules that govern how the microscopic interaction couples different subspaces.
The choice of cavity-selected subspace thus provides a route to engineering structured non-local interactions and to exploring many-body physics beyond standard local and collective descriptions.
\\
\\
\noindent \textbf{\textit{Cavity-Rydberg Spin Model.---}}
We consider an interacting system of $N$ spin-$\frac{1}{2}$s that is governed by the generalized XYZ Hamiltonian
\begin{equation}
    H
    =
    \lambda _x (S_x)^2 
    + 
    \lambda _y (S_y)^2
    +
    \sum _{i<j} J_{ij}
    S^z_i S^z_j
    .
    \label{eq:m_LMG_ising_hamiltonian}
\end{equation}
Here, $S_\alpha = \sum_j S_j^\alpha$ is the collective spin operator, with $S_\alpha\ket{M_\alpha} = M_\alpha \ket{M_\alpha}$ for $\alpha \in \{x,y,z\}$. The couplings $\lambda_x$ and $\lambda_y$ set the strengths of the collective interactions along the $x$- and $y$-axes, respectively. 
We take $J_{ij}$ such that the Ising term admits a finite Lieb-Robinson velocity \cite{Lieb_Robinson_CommMatPhys_1972_the, Hastings_Koma_CommMatPhys_2006_spectral, Nachtergaele_Sims_CommMatPhys_2006_lieb, Kuwahara_Saito_PRX_2020_strictly, Chen_RepProgPhys_2023_speed}, for example, $J_{ij}$ is finite range or exponentially decaying.

The short-range Ising interaction serves as a simple illustrative example, but the same operator-level reasoning extends to general local spin Hamiltonians.
We first analyze the symmetric limits of the all-to-all interaction, $\lambda _y = 0$ \cite{Marsh_PRL_2025_multimode, Lee_PRL_2004_first, Zhang_SciRep_2014_quantum} and $\lambda _x = \lambda _y$ \cite{Norcia_Science_2018_cavity, Davis_PRL_2019_photon, Mann_2025_squeezing}, and afterwards argue that our conclusions apply more broadly to the general case of $\lambda _x, \lambda _y \neq 0$ \cite{Sorensen_Molmer_PRA_2002_entangling, Dimer_PRA_2007_proposed, Larson_EPL_2010_circuit}.
\begin{figure}[ht]
    \centering
    \includegraphics[width=\linewidth]{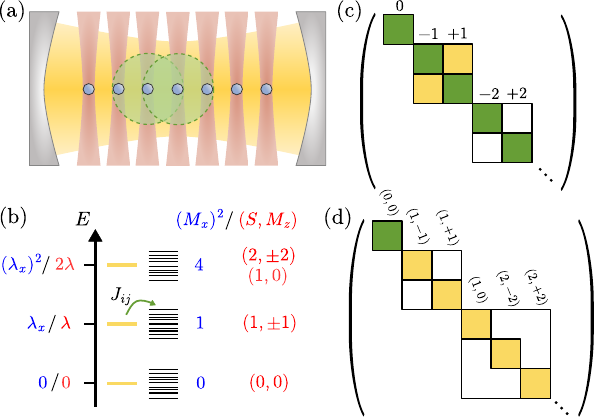}
    \caption{(a) A Rydberg-tweezer cavity interface: a delocalized cavity mode (yellow) mediates all-to-all interactions between atoms (blue) trapped by optical tweezers (red), while Rydberg dressing produces short-range interactions (green). 
    (b) We consider two paradigmatic collective models. In the first, $\lambda _y = 0$, the cavity creates degenerate subspaces labeled by $(M_x)^2$, while in the second, $\lambda _x = \lambda _y = \lambda$, the degenerate subspaces are labeled by $(S,M_z)$. 
    The values of $(M_x)^2$ and $(S,M_z)$ for the lowest-energy subspaces are respectively indicated on the right in blue and red, while their energies are indicated in the same color coding on the left. 
    The weak short-range Ising interactions $J_{ij}$ break the degeneracy of each subspace.
    (c) Projecting the Ising interaction into a subspace of fixed $Q = (M_x)^2$, according to the effective Hamiltonian (\ref{eq:m_effective_hamiltonian}), yields a block-diagonal structure.
    Sub-blocks are labeled by $M_x$ and shaded in green if the matrix elements of the Ising interaction correspond to a local effective interaction, while blocks shaded in yellow correspond to a non-local effective interaction. 
    (d) Block-diagonal structure of the effective Hamiltonian (\ref{eq:m_effective_hamiltonian}) based on energy for $Q = S(S+1) - (M_z)^2$. Sub-blocks are labeled by $(S, M_z)$ and the coloring code is the same as in (c).}
    \label{fig:main_setup_subspace}
\end{figure}
\,
\\
The family of Hamiltonians (\ref{eq:m_LMG_ising_hamiltonian}) can be implemented in cavity or circuit QED. 
Concretely, in a Rydberg tweezer-cavity interface \cite{Desantis_2026_realization} [Fig.~\hyperref[fig:main_setup_subspace]{\ref*{fig:main_setup_subspace}(a)}], two cavity-assisted Raman channels generate the all-to-all interactions \cite{Sorensen_Molmer_PRA_2002_entangling, Dimer_PRA_2007_proposed}, while Rydberg dressing techniques engineer short-range Ising interactions \cite{Zeiher_NatPhys_2016_many}. 

We work in the limit of strong all-to-all interactions such that transitions between different collective states are off-resonant. 
Concretely, at a fixed number of spins $N$, the limit $\lambda _{x,y}/J_{ij} \to \infty$ separates the spectrum into subspaces with distinct global spin quantum numbers, as visualized in Fig.~\hyperref[fig:main_setup_subspace]{\ref*{fig:main_setup_subspace}(b)}.
For example, in the case $\lambda _y = 0$ and $\lambda _x/J_{ij} \to \infty$, the all-to-all interaction is $\lambda _x (S_x)^2$ and each subspace carries the quantum number $(M_x)^2$. 
Although the short-range Ising interaction can couple different subspaces, these transitions are suppressed by the large gaps between subspaces. 
The effective dynamics then conserve the new quantum number, and the effective Hamiltonian within a subspace labeled by $Q$ is then obtained by projection \cite{Bravyi_Loss_AnnPhys_2011_schrieffer}:
\begin{equation}
    \mathbb{P}_Q
    H_\mathrm{eff}^Q
    =
    \mathbb{P}_Q
    \Big(
    \sum \nolimits_{i<j} J_{ij}
    S^z_i S^z_j
    \Big)
    \mathbb{P}_Q
    ,
    \quad
    [H_\mathrm{eff}^Q , \mathbb{P}_Q]
    =
    0
    ,
    \label{eq:m_effective_hamiltonian}
\end{equation}
where $\mathbb{P}_Q = \sum _\alpha \outprod{Q; \alpha}{Q;\alpha}$ is the projector into the $Q$-subspace and $\alpha$ is a multiplicity label.
The collective interaction therefore selects the accessible Hilbert space, while the projected Ising interaction governs the dynamics within it.

Although the projector $\mathbb{P}_Q$ originates from an all-to-all interaction, and is therefore intrinsically non-local, it need not produce a non-local effective Hamiltonian. Subspaces have been identified in which $H_{\mathrm{eff}}^Q$ inherits the spatial profile of $J_{ij}$ \cite{Santos_PRL_2016_cooperative, Celardo_PRB_2016_schielding, Mann_2025_squeezing}---a phenomenon we call subspace locality. 
We will show how the internal structure of the selected subspace either gives rise to subspace locality or structured non-locality.
\\
\\
\textbf{\textit{Product-State Subspace.---}} 
In the case where the all-to-all interaction is a collective transverse exchange $(S_x)^2$, $\lambda _y = 0$, the isolated subspaces are labeled by $(M_x)^2$ [Fig.~\hyperref[fig:main_setup_subspace]{\ref*{fig:main_setup_subspace}(b)}].

To derive the effective Hamiltonian $H_\mathrm{eff} ^{(M_x)^2}$, we first decompose each $(M_x)^2$-subspace into its two magnetization sectors, $\mathcal{H}_{(M_x)^2} = \mathcal{H}_{-M_x} \oplus \mathcal{H}_{+M_x}$; the $(M_x)^2 = 0$-subspace has the trivial decomposition $\mathcal{H}_{(M_x)^2 = 0} = \mathcal{H}_{M_x = 0}$.
Hence, the effective Hamiltonian $H_\mathrm{eff}^{(M_x)^2}$ has a $2 \times 2$ block structure [Fig.~\hyperref[fig:main_setup_subspace]{\ref*{fig:main_setup_subspace}(c)}] and each individual block is obtained by projecting the Ising interaction into the four sectors $(\pm M_x, \pm M_x)$ and $(\mp M_x, \pm M_x)$. 
Evaluating these blocks reduces to identifying the components of the Ising interaction that preserve the magnetization or change it by $2|M_x|$.

A suitable operator decomposition is provided by the natural basis of each magnetization sector: 
a basis state consists of $N/2 + M_x$ spins pointing along $+x$, and the remaining along $-x$. 
This defines a local product-state basis, because each basis state assigns every spin a definite $S^x_\ell$ eigenvalue, so that all pairs of sites $i$ and $j$ are disentangled. The Ising term can then be written in terms of local ladder-operator components with definite $\Delta M_x$:
\begin{equation}
    S^z_i S^z_j 
    = 
    \frac{1}{4} 
    ( 
        S^{x,+}_i S^{x,-}_j 
        + 
        S^{x,-}_i S^{x,+}_j  
        - 
        S^{x,+}_i S^{x,+}_j 
        - 
        S^{x,-}_i S^{x,-}_j 
    )
    .
    \label{eq:pss_ising_x-axis}
\end{equation}
Here, $S^{x,\pm}_\ell \ket{\mp}_\ell = \ket{\pm}_\ell$ is the spin ladder operator with respect to the $x$-quantization axis. The first two terms preserve $M_x$, while the latter two terms change it by $\pm 2$. 

After projecting Eq.~(\ref{eq:pss_ising_x-axis}) into the four magnetization sectors making up an $(M_x)^2$-subspace, the effective Hamiltonian is
\begin{equation}
\begin{split}
    H_\mathrm{eff} ^{(M_x)^2}
    =
    \sum _{i<j} 
    \frac{J_{ij}}{4}
    \Big\{
    &   
        S^{x,+}_i S^{x,-}_j 
        + 
        S^{x,-}_i S^{x,+}_j 
    \\
        &-
        \mathbb{P}_{M_x = 1}
        S^{x,+}_i S^{x,+}_j
        +
        \mathrm{H.c.}
    \Big\}
    .
\end{split}
\label{eq:m_effective_hamiltonian_mx2}
\end{equation} 
The first line on the right-hand side is present for all $(M_x)^2$-subspaces and describes an effective hopping along the $x$-quantization axis with the same interaction profile $J_{ij}$ as the microscopic Ising term \cite{Santos_PRL_2016_cooperative}. Hence, projection changes the operator content but not the interaction range, leading to subspace locality.

The $(M_x)^2 = 1$-subspace is exceptional, as shown by the extra term in the second line of Eq.~(\ref{eq:m_effective_hamiltonian_mx2}). The new term arises because the Ising interaction connects the $M_x = -1$ and $M_x = +1$ sectors, which does not change $(M_x)^2$. 
This term therefore produces an effective pair flip whose amplitude retains the local profile $J_{ij}$ but whose activation depends on the total magnetization $M_x$.
Thus, the same local spin configuration can evolve differently depending on spins arbitrarily far away. 

This non-locality can be exposed operationally through a simple communication protocol \cite{Xu_Swingle_PRXQuantum_2024_scrambling} (\textit{Appendix A}). 
Consider an open chain of $N$ spins, of which Alice controls the first two $(1,2)$ and Bob the last two $(N-1,N)$ spins. The initial state is chosen to be
\begin{equation*}
    \ket{\widetilde{\psi}}
    =
    \ket{ \! ++}_{1,2}
    \otimes
    \ket{M_x = 1}_{3, \dots, N-2}
    \otimes
    \ket{ \! - -}_{N-1,N}
    .
\end{equation*}
This state lies in the $(M_x)^2 = 1$-subspace and the bulk state $\ket{M_x = 1}_{3, \dots, N-2}$ can be any state with magnetization $M_x = +1$. 
Alice encodes the classical bit $1$ in the initial state by flipping her spins, $\ket{++}_{1,2} \to \ket{--}_{1,2}$, and the bit $0$ by doing nothing. 
The two choices place the full system in the $M_x = -1$ and $M_x = +1$ sectors, respectively.

To concentrate the conditioned local dynamics on Bob's spins, we choose the Ising profile to be nonzero only on his sites, $J_{ij} = -J\delta_{i,N-1} \delta_{j,N}$.
The effective Hamiltonian is thus
\begin{equation*}
    \widetilde{H}_\mathrm{eff}
    =
    \frac{J}{4}
    \big(
        \mathbb{P}_{M_x = 1}
        S^{x,+}_{N-1} S^{x,+}_N
        +
        \mathbb{P}_{M_x = -1}
        S^{x,-}_{N-1} S^{x,-}_N
    \big)
    .
\end{equation*} 
We omit the hopping term that acts on the sites $(N-1,N)$ because it annihilates the initial state and every state dynamically connected to it.

\begin{figure}
     \includegraphics[width=1.0\linewidth]{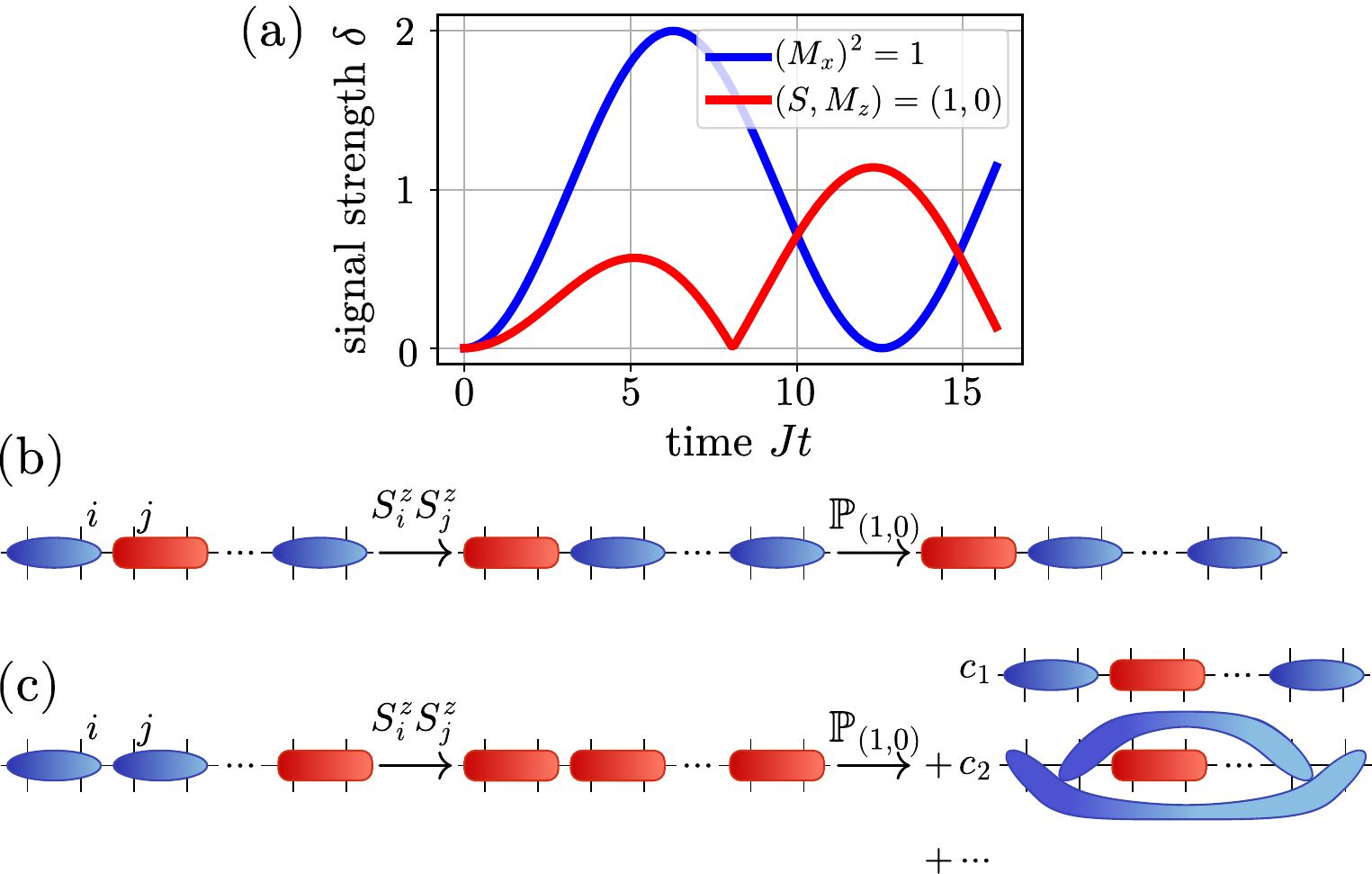}
     \caption{(a) Signal strength $\delta = | \langle O_\mathsf{B} (t) \rangle _0 - \langle O_\mathsf{B} (t) \rangle _1 |$ versus time $Jt$ for the product-state subspace protocol (blue) and the entangled-subspace protocol (red). 
     The indices $0$ and $1$ represent Alice's bit, and Bob's observable, $O_\mathsf{B}$, is $2 S^x_{N}$ (blue) or $(\mathbf{S}_{N-1} \cdot \mathbf{S}_{N})^2$ (red).
     A non-zero value of $\delta$ allows Bob to distinguish Alice's encoded bit. 
     Because $\delta$ is independent of system size $N$ and the separation between Alice and Bob grows with $N$, both protocols demonstrate distance-independent signaling.
     (b) \& (c) Action of a projected local Ising interaction on sites $i$ and $j$ for two valence-bond configurations in the $(S = 1, M_z = 0)$-subspace. 
     Blue ovals denote singlets and red rounded squares denote non-magnetic triplets. 
     Locally indistinguishable configurations evolve differently because the outcome depends on the remote triplet position, demonstrating that the projected interaction is non-local. 
     In (c), the projected state is expanded in an overcomplete valence-bond basis. 
     The dots denote the two remaining permutations of the triplet location.
     }
     \label{fig:main_communication}
\end{figure}

If Alice encodes the bit $1$, the state changes its total magnetization from $M_x = +1$ to $M_x = -1$, activating the conditioned pair flip on Bob's bond. 
His spins then undergo Rabi oscillations between the two configurations $\ket{--} _{N-1, N}$ and $\ket{++} _{N-1, N}$ with frequency $\Omega = J/2$.
By contrast, if Alice encodes the bit $0$, the pair flip is inactive and Bob's spins remain frozen. 
By performing a local measurement of the $x$-magnetization of spin $N$, Bob observes a measurable difference $\delta$ between the two cases [blue curve, Fig.~\hyperref[fig:main_communication]{\ref*{fig:main_communication}(a)}]. 
Hence, Alice can transmit a classical signal to Bob in a time that is independent of their separation.
Such distance-independent signaling is impossible under a short-range Hamiltonian, for which the Lieb-Robinson bound limits information propagation for every initial state \cite{Xu_Swingle_PRXQuantum_2024_scrambling}. 

The order-one local signature of this non-locality relies on the engineered initial state and interaction profile.
For a generic product state in the $M_x = -1$-sector and bulk Ising interactions, the pair-flip process $\ket{--} \to \ket{++}$ can instead occur on any eligible bond.
Although this distribution is expected to dilute the signal in any fixed local observable, the engineered protocol demonstrates that the effective Hamiltonian itself is non-local, but highly structured.

\,
\\
\noindent \textbf{\textit{Entangled Subspace.---}} 
The all-to-all interaction $\mathbf S^2-(S_z)^2$, $\lambda _x = \lambda _y$, selects a qualitatively different set of subspaces with intrinsic angular-momentum entanglement.
Because the Ising interaction conserves $M_z$, the projected dynamics decomposes into blocks labeled by $(S,M_z)$ even when several such blocks are energetically degenerate [Figs.~\hyperref[fig:main_setup_subspace]{\ref*{fig:main_setup_subspace}(b)} and \hyperref[fig:main_setup_subspace]{(d)}].
In contrast to the $(M_x)^2$-subspaces, a fixed $(S,M_z)$-subspace generically does not admit a local product-state basis. Projection therefore generically dresses local perturbations by global operators, producing non-local but structured dynamics.

The distinction between the subspaces originates from the non-trivial addition of angular momentum.
While two subsystems with magnetizations $M_{x,1}$ and $M_{x,2}$ combine to form a bigger system with well-defined magnetization, a product of subsystem states with spins $S_1$ and $S_2$ generally decomposes into a superposition of total-spin sectors $S^\prime$ satisfying $|S_1 - S_2| \leq S^\prime \leq S_1 + S_2$.
A definite total spin is therefore not specified by a single configuration of local quantum numbers.

A coupled basis state with fixed total spin $S$ can be written as a superposition of subsystem configurations with the appropriate Clebsch-Gordan coefficients:
\begin{equation*}
    \ket{S, M_z; a,b}
    =
    \sum _{m_z}
    c_{m_z}
    \ket{S_1, m_z; a}
    \otimes
    \ket{S_2, M_z - m_z; b}
    .
\end{equation*}
Here, $a$ and $b$ label multiplicities and $c_{m_z}$ abbreviates the Clebsch-Gordan coefficient $C_{S_1, m_z;S_2, M_z - m_z}^{S, M_z} = \inprod{S_1, m_z; S_2, M_z - m_z}{S, M_z}$.
For a generic basis state $\ket{S, M_z; a,b}$, more than one Clebsch-Gordan coefficient is non-zero, so the state is entangled across the bipartition. 
Thus, generic fixed-$(S,M_z)$ subspaces are intrinsically entangled and do not admit a product-state basis \cite{Li_PRX_2025_highly, Moharramipour_PRXQuantum_2024_symmetry}.

A comparison between two valence-bond configurations makes the resulting non-locality explicit.
Consider the projected Ising interaction $S_i^z S_j^z$ in the $(S = 1, M_z = 0)$-subspace.
This subspace is spanned by valence bond configurations where all atoms pair up into singlet bonds, $\ket{s} = \frac{1}{\sqrt{2}} (\ket{\! \uparrow \downarrow} - \ket{\! \downarrow \uparrow})$, except two atoms which pair into a triplet bond, $\ket{t} = \frac{1}{\sqrt{2}} (\ket{\! \uparrow \downarrow} + \ket{\! \downarrow \uparrow})$. 

In the first configuration [Fig.~\hyperref[fig:main_communication]{\ref*{fig:main_communication}(b)}] the site $i$ is part of a singlet and site $j$ part of a triplet, so the action of $S^z_i S^z_j$ just transforms the singlet into a triplet and the triplet into a singlet.
Hence, the Ising interaction simply moves the triplet bond, which preserves the total spin so the subsequent projection $\mathbb{P}_{(S = 1, M_z = 0)}$ acts trivially on this state.
In the second configuration [Fig.~\hyperref[fig:main_communication]{\ref*{fig:main_communication}(c)}] the sites $i$ and $j$ belong to two distinct singlets.
Acting with $S_i^z S_j^z$ converts both singlet bonds into triplets, which produces three triplet bonds that, by the rules of angular momentum addition, have components in the $S = 1$ and $S = 3$ subspaces.
The effect of projection $\mathbb{P}_{(S = 1, M_z = 0)}$ is now non-trivial as it removes the contribution in the $S = 3$ subspace, leaving behind the $S = 1$ component of the three triplets.
In the state remaining after projection, sites $i$ and $j$ are entangled with the spins where the original triplet was located, which may be arbitrarily far away.
While the two initial configurations differ globally, they have the same reduced density matrix on sites $i$ and $j$, $\rho _{ij}^\mathrm{red} \propto \mathds{1}_i \otimes \mathds{1}_j$, hence no local operator acting only on the sites $i$ and $j$ can distinguish them. 
Because the projected Ising interaction acts differently in the two cases, its effective action must be non-local.

To systematically describe this mechanism, we need the analogue of the $M_x$-ladder operator decomposition used in the product-state case [cf.~Eq.~(\ref{eq:pss_ising_x-axis})]. 
There, local operators were decomposed into local terms that changed $M_x$ by a definite amount. 
In a total spin subspace, however, no decomposition into local operators exists whose components change the total spin by a definite amount. 
Instead the appropriate local decomposition is formed by irreducible spherical tensors $T^{(k)}_q$ \cite{Sakurai_Napolitano_2017_modern, Varshalovich_1987_quantum} (\textit{Appendix B}): acting with $T^{(k)}_q$ on $\ket{S, M_z}$ produces a superposition of total spin states $S^\prime$ that obey $|S - k| \leq S^\prime \leq S + k$ and changes the magnetization by $q$.
The Ising interaction $S^z_i S^z_j$ then decomposes as
\begin{equation*}
    \frac{1}{3}
    \mathbf{S}_i \cdot \mathbf{S}_j
    +
    \Big(
        S^z_i S^z_j 
        - 
        \frac{1}{3}
        \mathbf{S}_i \cdot \mathbf{S}_j
    \Big)
    =
    -
    \sqrt{\frac{1}{3}}
    T^{(0)}_{0, ij}
    +
    \sqrt{\frac{2}{3}}
    T^{(2)}_{0, ij}
    .
\end{equation*}
The scalar component $T^{(0)}_{0, ij} \propto \frac{1}{3} \mathbf{S}_i \cdot \mathbf{S}_j$ preserves $S$ and remains local after projection. 
Therefore, any non-locality must originate from the projected quadrupolar component $T^{(2)}_{0, ij} \propto S^z_i S^z_j - \frac{1}{3} \mathbf{S}_i \cdot \mathbf{S}_j$.

The Wigner-Eckart theorem naturally exposes the structure of the non-local term and, within a fixed-$S$ sector, reads \cite{Sakurai_Napolitano_2017_modern}:
\begin{equation*}
    \bra{S, M_z^\prime; \alpha ^\prime}
    T^{(k)}_{q, ij} 
    \ket{S, M_z; \alpha}
    =
    \langle S; \alpha ^\prime
    \|
    \mathbf{T}^{(k)}_{ij}
    \|
    S; \alpha \rangle
    \,
    C^{S, M_z^\prime}_{S, M_z ; k, q}
    .
\end{equation*}
It states that the matrix element of a spherical tensor $T^{(k)}_{q, ij}$ factorizes into two pieces. 
The first factor is called the reduced matrix element and contains the microscopic structure of the operator and is independent of the orientation of the total spin. 
The second factor, the Clebsch-Gordan coefficient, describes how the tensor component changes the orientation of the total spin and is fixed entirely by angular-momentum addition.

This matrix-element factorization can be written at the operator level as the tensor product (\textit{Appendix C})
\begin{equation}
    \mathbb{P}_S
    T^{(k)}_{q,ij}
    \mathbb{P}_S
    =
    A^{(S,k)}(i,j)
    \otimes
    \tau^{(S, k)}_{q}
    .
    \label{eq:m_wigner_eckart_operator}
\end{equation}
The reduced operator $A^{(S,k)}(i,j)$ contains the microscopic and spatial dependence and acts on the multiplicity space, whereas $\tau ^{(S, k)}_{q}$ encodes the Clebsch-Gordan coefficients and acts on the angular-momentum multiplet $\{ \ket{S, M_z}\} _{M_z}$.

To obtain an effective operator for $T^{(2)}_{0,ij}$, it remains to find an expression for the reduced operator $A^{(S,2)}(i,j)$ in terms of spin operators in the full Hilbert space.
In \textit{Appendix C}, we show how to “invert” Eq.~(\ref{eq:m_wigner_eckart_operator}) by contracting both sides with the collective quadrupole tensor $\bm{\mathcal{T}}^{(2)}$ whose components are proportional to $\tau ^{(S, 2)}_{q}$ within a fixed-$S$ sector.
Combining the resulting quadrupole contribution with the scalar component gives the effective Hamiltonian in an $(S,M_z)$-subspace,
\begin{equation}
\begin{split}
    H_\mathrm{eff} ^{(S, M_z)}
    =
    &\sum _{i<j}
    J_{ij}
    \bigg\{   
        \frac{1}{3}
        \mathbf{S}_i \cdot \mathbf{S}_j
    \\
    &
    \hspace{15pt}
    +
        \frac{f_{S, M_z}}{2}
        \sum _{q = -2}^2
        (-1)^q
        T^{(2)}_{q, ij}
        \mathcal{T}^{(2)}_{-q}
        +
        \mathrm{H.c.}
    \bigg\}
    ,
\end{split}
\label{eq:ess_effective_hamiltonian_Smz}
\end{equation}
where $\frac{f_{S, M_z}}{2} = \frac{3 (M_z)^2 - S(S+1)}{S(S+1)(2S - 1) (2S + 3)}$ for $S > \frac{1}{2}$ and zero otherwise. 
The first term is a local Heisenberg interaction that directly inherits the Ising interaction profile $J_{ij}$. 
The second term couples the quadrupole of the sites $i$ and $j$ with strength $J_{ij}$ to the quadrupole of the entire system. 
Hence, projection changes both the local operator content and dresses local processes with a global operator.
The resulting dynamics is non-local but retains a local times global structure.

Subspace locality emerges when angular-momentum addition forces the quadrupolar contribution to vanish. This occurs when $f_{S, M_z} = 0$, equivalently when $3 (M_z)^2 - S (S+1) = 0$.
In these subspaces, the effective Hamiltonian $H_\mathrm{eff} ^{(S, M_z)}$ reduces to a local Heisenberg Hamiltonian, which relates to the Heisenberg spin liquids discussed in Ref.~\cite{Mann_2025_squeezing}. 
These exceptional subspaces are the many-body analogue of atomic selection rules: the Ising interaction can change the total spin, and the coupling strength between total spin states is governed by the rules of angular momentum addition.
\\
\\
The non-locality of $H_{\mathrm{eff}}^{(S,M_z)}$ (\ref{eq:ess_effective_hamiltonian_Smz}) can be exposed via a simple communication protocol \cite{Xu_Swingle_PRXQuantum_2024_scrambling} (\textit{Appendix A}).
It is based on the previous observation, visualized in Figs.~\hyperref[fig:main_communication]{\ref*{fig:main_communication}(b)} and \hyperref[fig:main_communication]{(c)}, that the action of a projected local Ising interaction depends on the position of the non-magnetic triplet. 
Consider an open chain in the $(S,M_z) = (1,0)$ subspace, where Alice controls sites $(2,3)$ and Bob sites $(N-1,N)$. We choose the initial state
\begin{equation*}
    \ket{\widetilde{\phi}}
    =
    \ket{t_{1,2}}
    \ket{s_{3,N-2}}
    \otimes
    \ket{S=0}_{4,\dots,N-3}
    \otimes
    \ket{s_{N-1,N}}
    ,
\end{equation*}
where $\ket{s_{ij}}$ and $\ket{t_{ij}}$ denote a singlet and non-magnetic triplet bond, respectively, and the bulk state can be any singlet. 
Alice encodes a classical bit by either locally exchanging the singlet and triplet character of the bonds $(1,2)$ and $(3,N-2)$, $\ket{t_{1,2}} \ket{s_{3,N-2}} \to \ket{s_{1,2}} \ket{t_{3,N-2}}$, or leaving the state unchanged.
Because a singlet and a non-magnetic triplet differ only by a relative phase, this exchange can be implemented by an operation supported entirely on Alice's sites. 
We take the Ising interaction to act only on the bond $(N-2, N-1)$ with strength $-J$. 
Alice's choice therefore determines whether the triplet bond overlaps with site $N-2$, causing Bob's sites to undergo different dynamics for the two encodings.
By measuring the total spin of his pair, Bob can distinguish Alice's choice after a time independent of their separation [red curve, Fig.~\hyperref[fig:main_communication]{\ref*{fig:main_communication}(a)}], demonstrating distance-independent signaling.
\\
\\
\noindent \textbf{\textit{Anisotropic Collective Interactions.---}}
We ask which side of the criterion governs the anisotropic all-to-all interaction $(S_x)^2 + \gamma(S_y)^2$.
The answer depends on whether the collective interaction energetically selects subspaces with or without a local product-state basis.
Although the total spin $S$ is always a good quantum number, the relevant distinction is whether the collective interaction energetically resolves different total-spin sectors. 
For $\gamma = 0$, the collective energy depends only on $(M_x)^2$, producing extensive degeneracies across different total-spin sectors, even before accounting for multiplicities. 
These large degeneracies allow the $(M_x)^2$-subspaces to be spanned by local product states, even though individual fixed-$S$ sectors are entangled \cite{Li_PRX_2025_highly}. 
For generic $\gamma \neq 0$, the interaction resolves the underlying total-spin structure and the systematic extensive degeneracy between different total-spin sectors is lifted, apart from accidental level crossings. 
Within each fixed-$S$ sector, any remaining degeneracy is at most twofold (\textit{Appendix D}) and the energetically selected subspaces therefore inherit the entanglement enforced by angular-momentum addition. 
Hence, the product-state structure of the $(S_x)^2$ limit is fine-tuned: generic anisotropy energetically selects subspaces that do not admit a product-state basis and hence, in general, structured non-local effective dynamics.
\\
\\
\textbf{\textit{Outlook.---}} 
We have shown that, up to angular-momentum selection rules, the projected dynamics under dominant collective coupling is organized by whether the selected subspace admits a local product-state basis.
Global projection can generate new short-range interactions or dress local processes with global operators.

This distinction extends beyond cavity-QED spin models. 
Product-state subspaces allow local perturbations to be decomposed into components with definite changes of the local quantum numbers, as in fixed particle-number sectors of bosons or fermions. 
By contrast, non-Abelian global quantum numbers such as those of $\mathrm{SU}(N)$ require non-trivial recoupling of local degrees of freedom, which generically dresses local processes with global tensor operators.

Structured non-local interactions have implications for both quantum information and many-body physics.
Local times global interactions could enable multiqubit-controlled gates, or map global quantum information onto local observables.
These interactions could also produce low-energy many-body behavior not captured by descriptions developed for local Hamiltonians.
A concrete example is the cavity-induced quantum spin liquid of Ref.~\cite{Mann_2025_squeezing}: its ground state is described by a local Heisenberg Hamiltonian, whereas spinful excitations are generically governed by structured non-local interactions [Eq.~(\ref{eq:ess_effective_hamiltonian_Smz})].
Together, these results define a class of non-local many-body dynamics that may support phenomena without purely local or collective counterparts.

\section*{Acknowledgements}
\begin{acknowledgements}
We thank Jamir Marino, Pablo Sala, Kaden Hazzard, Lorenzo Rossi and Hossein Hosseinabadi for stimulating discussions, and M.A.O.~thanks Frank Pollmann for an illuminating talk about Ref.~\cite{Li_PRX_2025_highly} at the conference 'Quantum Simulation with Engineered Dissipation' in Obergurgl. We are grateful to Manuele Landini for pointing us to a valuable reference. 

M.A.O.~was supported by a Marie Skłodowska-Curie Actions-COFUND PhD fellowship (No.~101081441). 
B.J.~was supported by the MSCA Postdoctoral Fellowship QUINTO (No.101145886). 
D.E.C. acknowledges support from the European Union, under European Research Council grant NEWSPIN (No 101002107), EIC Pathfinder Grant PANDA (No 101115420); the Government of Spain (Project PID2024-158422NB-I00 funded by MICIU/AEI/10.13039/501100011033 FEDER and Severo Ochoa Grant CEX2024-001490-S [MICIU/AEI/10.13039/501100011033]); QuantERA II project QuSiED, co-funded by the European Union Horizon 2020 research and innovation programme (No 101017733) and the Government of Spain (European Union NextGenerationEU/PRTR PCI2022-132945 funded by MCIN/AEI/10.13039/501100011033); Generalitat de Catalunya (CERCA program and AGAUR Project No. 2021 SGR 01442); Fundacio Cellex, and Fundació Mir-Puig.
C.-R.M.~was supported by the MSCA Postdoctoral Fellowship ATOMAG (No. 101068503). 
The authors also acknowledge the hospitality of the Kavli Institute for Theoretical Physics (NSF PHY-2309135).

\textit{Use of AI Tools.---} We acknowledge the use of ChatGPT-5 to ChatGPT-5.6 for discussion, review and editing. While we had proved that the effective Hamiltonian in an $(S, M_z)$-subspace must be either the local Heisenberg model or non-local, we did not derive an explicit expression for it. ChatGPT 5.5 then proposed to perform a tensor contraction to isolate $A^{(S,k)}(i,j)$ which led to Eq.~(\ref{eq:ess_effective_hamiltonian_Smz}).

\end{acknowledgements}

\bibliography{references}

\newpage

\appendix

\section{End Matter}

\noindent\textit{Appendix A: Set-up of the Classical Communication Protocol.---}
The discussion of the communication protocol follows Ref.~\cite{Xu_Swingle_PRXQuantum_2024_scrambling}.

Consider an open chain of $N$ spin-$\frac{1}{2}$s, where Alice controls the first few spins and Bob the last few, so that their separation scales linearly with $N$. Starting from the initial state $\ket{\psi}$, Alice encodes a classical bit $a \in \{ 0,1 \}$ by either doing nothing ($a=0$) or applying a local unitary $u_\mathsf{A}$ ($a=1$). The system then evolves under $U(t) = \e^{-\i Ht}$, after which Bob measures a local observable $O_\mathsf{B}$ with the aim of identifying Alice's classical bit.

For $a = 0$, Bob's measurement becomes
\begin{equation*}
    \langle O_\mathsf{B} \rangle  _0
    =
    \bra{\psi} U^\dagger (t) O_\mathsf{B} U(t) \ket{\psi}
    =
    \bra{\psi (t)} O_\mathsf{B} \ket{\psi (t)}
    ,
\end{equation*}
where $\ket{\psi (t)} = U(t) \ket{\psi}$. On the other hand, for $a = 1$, the initial state is $u_\mathsf{A} \ket{\psi}$ and therefore
\begin{align*}
    \langle O_\mathsf{B} \rangle  _1
    &=
    \big(
        \bra{\psi} u_\mathsf{A}^\dagger 
    \big)
        U^\dagger (t) O_\mathsf{B} U(t) 
    \big(
        u_\mathsf{A} 
        \ket{\psi}
    \big)
    \\
    &=
    \bra{\psi (t)} u_\mathsf{A}^\dagger (-t) O_\mathsf{B} u_\mathsf{A} (-t) \ket{\psi (t)}
    ,
\end{align*}
with $u_\mathsf{A} (-t) = U(t) u_\mathsf{A} U^\dagger (t)$.

We quantify Bob's ability to distinguish the two encodings by the signal strength
\begin{equation*}
    \delta (t)
    =
    \big|
        \langle O_\mathsf{B} \rangle  _0
        -
        \langle O_\mathsf{B} \rangle  _1
    \big|
    .
\end{equation*}
The signal is bounded by the spreading of Alice's initially local unitary to Bob's sites \cite{Xu_Swingle_PRXQuantum_2024_scrambling}:
\begin{equation*}
    \delta ^2 (t)
    \leq
    \big \| 
        [u_\mathsf{A} (-t), O_\mathsf{B}]
    \big \| ^2 _\infty
    \leq
    c \,
    \e ^{- \mu [ \mathsf{d}(\mathsf{A}, \mathsf{B}) - v_\mathrm{LR} t ]}
    .
\end{equation*}
Here, $\| \cdot \|_\infty$ is the operator norm, $c$ and $\mu$ are finite constants, and $\mathsf{d}(\mathsf{A}, \mathsf{B})$ is the distance between Alice and Bob. For a short-range Hamiltonian, $v_\mathrm{LR}$ is finite, so Bob's signal is exponentially suppressed until a time that grows with their separation. This is a consequence of the Lieb-Robinson bound \cite{Lieb_Robinson_CommMatPhys_1972_the, Hastings_Koma_CommMatPhys_2006_spectral, Nachtergaele_Sims_CommMatPhys_2006_lieb, Chen_RepProgPhys_2023_speed}, which bounds the propagation of information in non-relativistic, local quantum systems. 

The protocol therefore provides an operational diagnostic of non-local dynamics. 
An order-one signal that appears at a time independent of $\mathsf{d}(\mathsf{A},\mathsf{B})$ demonstrates distance-independent signaling, which cannot occur under any short-range Hamiltonian.
\\
\\
\noindent\textit{Appendix B: Definition of Spherical Tensor Operators.---}
We fix the normalization convention for the spherical tensors used in this work by defining the single-site rank-$1$ spherical components as
\begin{equation*}
    T^{(1)}_{0,j}
    =
    S^z_j
    \quad
    \text{and}
    \quad
    T^{(1)}_{\pm 1,j}
    =
    \mp 
    \frac{1}{\sqrt{2}}
    S^\pm_j
    ,
\end{equation*}
and constructing higher-rank tensors through the Clebsch-Gordan coupling \cite{Sakurai_Napolitano_2017_modern},
\begin{equation}
    \Big[ 
        \mathbf{A}^{(k_1)}
        \otimes 
        \mathbf{B}^{(k_2)}
    \Big]
    _q ^{(k)}
    =
    \sum _{q_1, q_2}
    C^{k,q}_{k_1, q_1; k_2, q_2}
    A^{(k_1)}_{q_1}
    B^{(k_2)}_{q_2}
    ,
    \label{eq:a_product_spherical_tensors}
\end{equation}
where $C^{k,q}_{k_1, q_1; k_2, q_2}$ is the Clebsch-Gordan coefficient in the Condon-Shortley phase convention.

In particular, using $T^{(2)}_{q,ij} = \big[ \mathbf{S}^{(1)}_i \otimes \mathbf{S}^{(1)}_j \big] _q ^{(2)}$ and $\mathcal{T}^{(2)}_{q} = \big[ \mathbf{S}^{(1)} \otimes \mathbf{S}^{(1)} \big] _q ^{(2)}$, we have
\begin{align*}
    T^{(2)}_{2,ij}
    =
    \frac{S^+_i S^+_j}{2}
    \quad
    &\text{and}
    \quad
    \mathcal{T}^{(2)}_{2}
    =
    \frac{(S^+)^2}{2}
    ,
    \\
    T^{(2)}_{1,ij}
    =
    -
    \frac{S^+_i S^z_j + S^z_i S^+_j}{2}
    \quad
    &\text{and}
    \quad
    \mathcal{T}^{(2)}_{1}
    =
    -
    \frac{S^+ S^z + S^z S^+}{2}
    ,
    \\
    T^{(2)}_{0,ij}
    =
    \frac{3 S^z_i S^z_j - \mathbf{S}_i \cdot \mathbf{S}_j}{\sqrt{6}}
    \quad
    &\text{and}
    \quad
    \mathcal{T}^{(2)}_{0}
    =
    \frac{3 (S_z)^2 - \mathbf{S}^2}{\sqrt{6}}
    ,
\end{align*}
respectively, where the remaining components are obtained by the relation $\big( T^{(2)}_{q} \big)^\dagger = (-1)^q T^{(2)}_{-q}$.
\\
\\
\noindent\textit{Appendix C: Derivation of the Effective Hamiltonian in an $(S,M_z)$-Subspace.---}
Here, we first recast the Wigner-Eckart theorem as the operator identity in Eq.~(\ref{eq:m_wigner_eckart_operator}) and then use this identity to derive $H_\mathrm{eff}^{(S,M_z)}$ (\ref{eq:ess_effective_hamiltonian_Smz}). 

Within a fixed-$S$ sector, the Hilbert space can be written as $\mathcal H_S \simeq \mathcal{M}_S \otimes \mathcal{V}_S$, where $\mathcal M_S$ carries the multiplicity label $\alpha$ and $\mathcal V_S$ is spanned by the angular-momentum states $\{ \ket{S,M_z} \}_{M_z}$. 
A basis state can therefore be written as the tensor product $\ket{S, M_z; \alpha} = \ket{\alpha} \otimes \ket{S, M_z}$.

For the $q$th component of a rank-$k$ irreducible spherical tensor operator, $T^{(k)}_{q,ij}$, the Wigner-Eckart theorem within a fixed-$S$ sector is \cite{Sakurai_Napolitano_2017_modern}
\begin{equation*}
    \bra{S,M_z^\prime;\alpha^\prime}
    T^{(k)}_{q,ij}
    \ket{S,M_z;\alpha}
    =
    \langle S;\alpha^\prime
    \|
    \mathbf T^{(k)}_{ij}
    \|
    S;\alpha\rangle
    \,
    C^{S,M_z^\prime}_{S,M_z;k,q}
    .
\end{equation*}
The reduced matrix element depends only on the multiplicity labels and the total spin, and the Clebsch-Gordan coefficient contains the $M_z$, $M_z^\prime$ and $q$ dependence.
To promote the Wigner-Eckart theorem to an operator statement, we define the reduced operator $A^{(S,k)}(i,j)$ through 
\begin{equation*}
    \bra{\alpha^\prime} A^{(S,k)}(i,j) \ket{\alpha} = \langle S;\alpha^\prime \| \mathbf T^{(k)}_{ij} \| S;\alpha\rangle
    ,
\end{equation*}
and the angular-momentum operator $\tau_q^{(S,k)}$ through 
\begin{equation*}
    \bra{S,M_z^\prime} \tau_q^{(S,k)} \ket{S,M_z} = C^{S,M_z^\prime}_{S,M_z;k,q}
    .
\end{equation*}
The factorization in the Wigner-Eckart theorem on the level of matrix elements then becomes a tensor product of operators on the two Hilbert spaces, $\mathcal{M}_S$ and $\mathcal{V}_S$, and the theorem can be written as the operator relation $\mathbb{P}_S T^{(k)}_{q,ij} \mathbb{P}_S = A^{(S,k)}(i,j) \otimes \tau_q^{(S,k)}$, which is Eq.~(\ref{eq:m_wigner_eckart_operator}) of the main text.

To derive $H_\mathrm{eff}^{(S,M_z)}$ (\ref{eq:ess_effective_hamiltonian_Smz}), we express the reduced operator $A^{(S,k)}(i,j)$ in terms of physical spin operators that live in the full Hilbert space. For this purpose, consider a collective spherical tensor $\mathcal{T}_q^{(k)}$ constructed entirely from the collective spin operators such that $[\mathcal{T}_q^{(k)}, \mathbb{P}_S] = 0$. Since collective spin operators do not act on the multiplicity labels, the Wigner-Eckart theorem gives
\begin{equation}
    \mathbb{P}_S
    \mathcal{T}_q^{(k)}
    \mathbb{P}_S
    =
    g_{S,k}
    \,
    \mathds{1}_{\mathcal M_S}
    \otimes
    \tau _q^{(S,k)},
    \label{eq:a_global_tensor_we}
\end{equation}
where $g_{S,k}$ is a number fixed by $S$, $k$, and the normalization of $\mathcal{T}_q^{(k)}$ (\textit{Appendix B}). Thus, the local tensor $T^{(k)}_{q,ij}$ and the collective tensor $\mathcal{T}_q^{(k)}$ have the same angular-momentum dependence inside a fixed-$S$ sector, while only the former contains the bond and multiplicity dependence.

Since the local and collective tensors have the same angular-momentum dependence, we can eliminate this common dependence and isolate $A^{(S,k)}(i,j)$ by coupling them to a rotational scalar. According to the tensor-product rule in Eq.~(\ref{eq:a_product_spherical_tensors}), the rank-zero coupling of two rank-$k$ tensors $\mathbf{A}^{(k)}$ and $\mathbf{B}^{(k)}$ is proportional to $\sum _{q=-k}^{k} (-1)^q A^{(k)}_{q} B^{(k)}_{-q}$. We therefore define the Hermitian contraction map
\begin{equation*}
    \frac{1}{2} 
    \sum _{q=-k}^{k} 
    (-1)^q 
    \Big[ 
        (\cdot ) 
        \, 
        \mathcal{T}^{(k)}_{-q} 
        + 
        \mathcal{T}^{(k)}_{-q} 
        \, 
        (\cdot )
    \Big]
    ,
\end{equation*}
such that the left-hand side of Eq.~(\ref{eq:m_wigner_eckart_operator}) becomes the Hermitian scalar operator
\begin{equation}
    K^{(k)}_{ij}
    =
    \frac{1}{2}
    \sum_{q=-k}^{k}
    (-1)^q
    \left[
    T^{(k)}_{q,ij}
    \mathcal{T}^{(k)}_{-q}
    +
    \mathcal{T}^{(k)}_{-q}
    T^{(k)}_{q,ij}
    \right]
    .
    \label{eq:a_tensor_contraction_local_global}
\end{equation}
Since $K^{(k)}_{ij}$ is a rotational scalar, it commutes with $\mathbf{S}^2$ and hence with $\mathbb{P}_S$.
The angular-momentum part of the right-hand side then is a rotational scalar, so that Schur's lemma implies that it is proportional to the identity on $\mathcal{V}_S$,
\begin{equation}
    \mathbb{P}_S
    \frac{1}{2}
    \sum_{q=-k}^{k}
    (-1)^q
    \left[
    \mathcal{T}^{(k)}_q
    \mathcal{T}^{(k)}_{-q}
    +
    \mathcal{T}^{(k)}_{-q}
    \mathcal{T}^{(k)}_q
    \right]
    \mathbb{P}_S
    =
    \lambda_{S,k}\mathbb{P}_S
    .
    \label{eq:a_tensor_contraction_global_global}
\end{equation}
Thus, the scalar operator $K^{(k)}_{ij}$ represents the reduced operator $A^{(S,k)}(i,j) $ within the fixed-$S$ sector:
\begin{equation*}
    \mathbb{P}_S 
    K^{(k)}_{ij} 
    \mathbb{P}_S 
    = 
    \frac{\lambda_{S,k}}{g_{S,k}} 
    A^{(S,k)}(i,j) 
    \otimes 
    \mathds 1_{\mathcal V_S}
    .
\end{equation*}
Plugging this operator expression for $A^{(S,k)}(i,j)$ and Eq.~(\ref{eq:a_global_tensor_we}) in the operator identity Eq.~(\ref{eq:m_wigner_eckart_operator}) yields
\begin{equation*}
    \mathbb{P}_S
    T^{(k)}_{q,ij}
    \mathbb{P}_S
    =
    \frac{1}{\lambda _{S,k}}
    \mathbb{P}_S
    K^{(k)}_{ij}
    \mathbb{P}_S
    \mathcal{T}_q^{(k)}
    \mathbb{P}_S
    .
\end{equation*}

For the Ising interaction, the non-scalar component has $k=2$ and $q=0$. Projecting further onto fixed $M_z$ replaces the remaining collective tensor by its diagonal matrix element,
\begin{equation*}
    \mathbb{P}_{S, M_z}
    \mathcal{T} ^{(2)}_0
    \mathbb{P}_{S, M_z}
    =
    \frac{3 M_z^2 - S(S+1)}{\sqrt{6}}
    \mathbb{P}_{S, M_z}
    .
\end{equation*}
With the normalization of \textit{Appendix B}, the collective-tensor self-contraction gives $\lambda _{S,2} = S(S+1)(2S-1)(2S+3)/6$ for $S > \frac{1}{2}$. 
For $S \leq 1/2$, the triangle rule $|S-2| \leq S^\prime \leq S + 2$ forbids a rank-$2$ tensor from acting within the same total-spin sector, so the quadrupolar contribution vanishes. 
Combining this quadrupolar contribution with the scalar component then gives Eq.~(\ref{eq:ess_effective_hamiltonian_Smz}).
\\
\\
\noindent\textit{Appendix D: Twofold Degeneracy for Anisotropic Collective Interactions.---}
Here, we show that, within each irreducible spin-$S$ representation $\mathcal{V}_S$, every eigenvalue of the anisotropic collective Hamiltonian has multiplicity at most two. 

Consider
\begin{align*}
    H_\mathrm{coll}
    &=
    \lambda _x 
    (S_x)^2 
    + 
    \lambda _y 
    (S_y)^2
    \\
    &=
    \lambda _+
    \Big[
        \mathbf{S}^2 
        - 
        (S_z)^2
    \Big]
    +
    \frac{\lambda _-}{2}
    \Big[
        (S_+)^2 
        + 
        (S_-)^2
    \Big]
\end{align*}
with $\lambda _\pm = (\lambda _x \pm \lambda _y)/2$.
The first term in the second line preserves $M_z$, while the second term changes it by $\pm 2$. 
Since $H_\mathrm{coll}$ is constructed from collective-spin operators, it commutes with $\mathbf{S}^2$. 
Moreover, since it changes $M_z$ by zero or $\pm 2$, it also commutes with the parity operator $\mathrm{e} ^{\mathrm{i} \pi S_z}$. 
Because $[\mathbf{S}^2, \mathrm{e} ^{\mathrm{i} \pi S_z}] = 0$, we may diagonalize $H_\mathrm{coll}$ independently within each $(S, \sigma)$-sector, where $\sigma$ is the parity eigenvalue.

Within a fixed parity sector, we order the basis states by decreasing $M_z$ in steps of two. Then, $H_{\mathrm{coll}}$ is represented within an $(S, \sigma)$-sector by the real symmetric tridiagonal matrix,
\begin{equation*}
    \begin{pmatrix}
        a^{S}_{M_z^\mathrm{max}} & b^{S}_{M_z^\mathrm{max}-2} & & & \\
         b^{S}_{M_z^\mathrm{max}-2} & a^{S}_{M_z^\mathrm{max}-2} &  b^{S}_{M_z^\mathrm{max}-4} & &  \\
        &  b^{S}_{M_z^\mathrm{max}-4} & \ddots & \ddots  \\
        & & \ddots & \ddots &  b^{S}_{M_z^\mathrm{min}}  \\
        & & & b^{S}_{M_z^\mathrm{min}} & a^{S}_{M_z^\mathrm{min}} \\
    \end{pmatrix}
    ,
\end{equation*}
where $M_z^{\max}$ and $M_z^{\min}$ denote the largest and smallest allowed magnetizations in that $(S, \sigma)$-sector.
The diagonal elements are
\begin{align*}
    a^S_{M_z}
    &=
    \lambda _+
    \big(
        S(S+1) - (M_z)^2
    \big)
\end{align*}
while the upper and lower diagonal elements are given by
\begin{align*}
    b^S_{M_z}
    &= 
    \frac{\lambda _-}{2}
    \sqrt{S(S+1) - (M_z + 1) (M_z + 2)}
    \\
    & \hspace{60pt}
    \times
    \sqrt{S(S+1) - M_z (M_z + 1)}
    .
\end{align*}
For $\lambda _- \neq 0$, every off-diagonal entry is non-zero, so each $(S,\sigma)$-block is an irreducible real symmetric tridiagonal matrix. 
Such matrices have a simple spectrum \cite{Gantmacher_Krein_2002_Oscillation}, so every eigenvalue is non-degenerate within a fixed $(S,\sigma)$-sector. 
At fixed $S$, there are only two parity sectors so that $H_\mathrm{coll}$ is at most twofold degenerate.
For $\lambda _- = 0$ the Hamiltonian $H_{\mathrm{coll}}$ is already diagonal in the $\{ \ket{S, M_z} \}$ basis and the twofold degeneracy within a fixed-$S$ sector arises from the energy invariance under $M_z \leftrightarrow - M_z$.

\end{document}